\documentclass[aip,reprint,twocolumn
 reprint,%
 amssymb, amsmath,%
 aip,cha,%
]{revtex4-1}

\usepackage{geometry}                
\usepackage{graphicx}
\usepackage{amssymb}
\usepackage{epstopdf}
\usepackage{hyperref}
\usepackage{fancyvrb}
\usepackage{listings}
\usepackage{cancel}
\begin{document}

\newcommand{\ps}{\begin{lstlisting}[frame=single,columns=fullflexible,basicstyle=\small\ttfamily]} 
\lstset{language=Prolog,basicstyle=\small}

\begin{center}
{\Large{\bf Teaching and learning mathematics with Prolog}}\\
{\sl Tom Bensky, Cal Poly San Luis Obispo, tbensky@calpoly.edu}\\
\today 
\end{center}

\section*{Abstract}
Procedural computer languages have long been used in many aspects of mathematics pedagogy.  In this work, we examine the use of Prolog, a {\sl declarative}  language for the same purpose.  We find the {\sl facts+rules} aspect of Prolog to be a novel platform for developing coding lessons to supplement the learning of mathematics.  Specific examples are presented.

\section{Introduction}
We have all likely used a computer to enhance our understanding of mathematics in some form or another. For mathematics pedagogy, there is no shortage of books on doing so.~\cite{engel,cheney,becker,devaney,guggenheimer,wolframbook,conrad}  Websites are also sources of mathematical lessons on coding.~\cite{websites}

Procedural coding languages are the familiar ones, like Python, C, Lua, Pascal, BASIC, or Javascript. Using these languages involve solving a problem with code, using a sequence of variables, loops, functions, that collectively implement some step-by-step logic (i.e. the ``procedure'') to solve the problem.  

This article suggests that coding in Prolog, a {\sl declarative language} offers some novel opportunities to study mathematics, not made available with procedural languages or computer algebra systems.

\section{Prolog: A brief introduction}

Coding in Prolog~\cite{clocksin,bratko,covington,bos} involves an entirely different thought process than needed by a procedural language. Instead of providing a step-by-step procedure, one provides two sets of information about the problem at hand: facts and rules. Prolog is then given a goal to try and satisfy, which sets it off finding combinations of the facts that follow the rules. Such combinations will comprise the ``solution'' to the stated goal.  As an example, suppose one is designing a travel-planning application in Prolog. 

{\sl The facts} would be cities on a map that are connected by roads, and what daily traffic patterns typically look like on the roads.  Facts are usually fairly straightforward to come by. In this case, the road and traffic data could easily be found on the Internet or even based on one's personal experience.  Facts are usually quantities one knows or can easily find or identify and are simply thought of as true.

{\sl The rules} about the problem are more difficult to formulate, and are more specific to the problem itself. Rules express in general terms, some relationship one would like to find from the facts. In the travel application, the rules would involve distance and traffic considerations when traveling from city A to city B, and some formulation that would minimize travel time. The rules would be very general, free of any specifics about either A or B, and readily applicable to the facts provided. A travel rule may very well include concerns for time of travel, speeds, distances, etc. This rule is to be applied to the facts to find for example, an optimal list of cities to visit.

Prolog is thus known as a {\sl declarative language} since one does not supply step-by-step instructions on how to solve a problem. Facts and rules are declared, and this is all Prolog needs to search for a solution.   The eventual outcome (or solution) will be all combinations of the facts that are consistent with the rules.  How does this work? 

Internal to Prolog is a run-time engine based around a well-implemented search algorithm. The algorithm reads in the facts and rules, and works to find combinations of facts that are consistent with the rules, automatically adjusting to the search landscape as it runs.  It will for example, abandon search paths that have no possibility of succeeding, while also keeping track of areas where a search looks more promising. We emphasize that one does not code the search functionality; this is built into the Prolog run-time engine.  (Rules are sometimes called ``clauses'' in Prolog, and will be used somewhat interchangeably here.)

{\sl The goal} in Prolog launches a search of the facts+rules in an attempt to satisfy the goal.  A goal is usually a statement or sequence of statements that guides Prolog in examaining the rules, to drive some problem specific logic.  Again in the travel application, after the city and road facts are set, followed by some travel rules, a goal may have the form of {\tt go(paris,rome).}, where {\tt go} is some top-level clause that starts the Prolog search. In very general terms, this may resemble something like {\tt go(A,B) :- road\_exists(A,B), low\_traffic(A,B), ...} and on, as required.

In the travel application, it hopefully seems plausible that that a shortest route between cities could in principle be found by looking at facts about roads and traffic. How the search should proceed however is not so intuitive. But {\sl Prolog handles the search aspects}, so one only need to focus on writing rules and presenting facts. This is a huge burden lifted off of the programmer, and allows for sole focus on describing the problem at hand via the facts and rules. In this regard, code is Prolog tends to be much shorter than an equivalent solution implemented in a procedural language.  Also, since the facts and rules are so plainly seen, coding features of the problem being solved stay in the forefront. Simple changes to the code are also usually straightforward to implement.~\cite{van_hetenryck}

\section{Studying Mathematics with Prolog}

We suggest that the facts, rules, and search characteristics of Prolog are ideal for studying mathematics. Why?  Because, it's hard to think of anything more fact and rule-based than mathematics. Facts here would be the statement of the problem.  The rules would be {\sl the rules} of mathematics itself, which are mostly what is taught in math classes. As in the travel application, facts for a math problem are again relatively easy to come by, with the rules being more difficult.   We assert here that the formulation of the rules into mathematics into Prolog is very natural, and is the supplemental exercises of the study of mathematics that we propose.

\subsection{Example: Adding two fractions}

As an example, think of coding a tutorial that shows how to add two fractions.~\cite{cbm}  The facts are plainly seen: the numerators and denominators of two fractions. Now what rules apply to add them properly?  This is a bit more difficult, because we know for instance that 1/4+1/2 is not 2/6. It would be incorrect to supply a rule that said ``The sum of two fractions is found by adding the numerators, then adding the denominators.'' A correct rule would say ``First look at the denominators.  If they are the same then the denominator is also used in the answer, and add the numerators to find the answer's numerator.'' Another rule is needed if the denominators are different.

Facts can be put in almost immediately and is a rather fun and inventive process, owing to Prolog's flexibility as a coding language.  Code representing fractions for example could be something like {\tt fract(1,4)}  or {\tt fract(1,2)}, to express the facts about the two fractions, one $1/2$ and the other $1/4$. We note that {\tt fract} is not a functor built into the Prolog language. It is of our own convenient making, that Prolog happily accepts.  Here {\tt fract} takes two parameters, a numerator and a denominator.

The rules however require more care. Independent of any coding, one must first be comfortable with all of the mathematical rules for adding fractions,  then must adapt them into Prolog code. Here's a rule to add two fractions that have the same denominator:

\ps
add(fract(N1,D1),fract(N2,D2),fract(N,D)) :- 
     D1 == D2, D = D1, N is N1 + N2.
\end{lstlisting}

\noindent The clause (or rule) called {\tt add} takes in two parameters, {\tt fract(N1,D1)} and {\tt fract(N2,D2)}.  You can see the use of the facts of the {\tt fract} functor. On entry, we expect {\tt N1, D1, N2} and  {\tt D2} to be instantiated (that is, equal to some numbers coming in). Prolog's job in the search as it uses this rule, is to find what {\tt N} and {\tt D} (the numerator and denominator of the fraction) will result from the sum of the two initial fractions, that is consistent with the {\tt add} rule.

When writing code, clauses are usually quietly translated into one's native tongue (here English). Prolog can be difficult to translate and perspectives on reading it are available.~\cite{pop_reading} (Quick Prolog review: {\tt :-} means ``if,'' {\tt ,} means ``and,'' {\tt ;} means ``or,'' variables are in uppercase, everything else must be in lowercase, and the period ({\tt .}) terminates facts and rules.) 

The so called ``procedural reading'' of the clause {\tt add} would be ``add can successfully add together $N_1/D_1$ and $N_2/D_2$ into $N/D$ IF the two ingoing denominators are equal AND $D$ can be instantiated to one of them (here $D_1$) AND $N$ can be instantiated to the sum of the two incoming numerators.''  Procedural readings often go from the clause head to the body of clause, in the order it is shown. In other words, the clause will succeed if each part of it succeeds in order they appear.

A ``declarative reading,'' which focuses on relationships between objects in the clause would be ``if we can set the sum numerator to the sum of the ingoing numerators and set the sum denominator to one of the ingoing denominators, and find that the two denominators equal, then adding $N_1/D_1$ and $N_2/D_2$ will result in $N/D$.'' Declarative readings go from the clause body back to its head as in: ``if each part of the body can be made to succeed, then the whole clause can succeed.'' The body is often even interpreted entirely in reverse order, from the period to the clause head.

These readings can be satisfying intellectual challenges, even more so with the backdrop of mathematics.

Prolog is excellent at pattern matching, so {\tt add} can also be simplified as

\ps
add(fract(N1,D),fract(N2,D),fract(N,D)) :- 
     N is N1 + N2.
\end{lstlisting}

\noindent Here you'll see how the denominator {\tt D} is placed in all positions it must be, for adding two fractions that have a common denominator: as the denominator of the two incoming fractions, and as the denominator of the resulting fraction.  The body now only needs to evaluate {\tt N} into the sum of {\tt N1 + N2}. This will also work:

\ps
add(fract(N1,D),fract(N2,D),fract(N1+N2,D)).
\end{lstlisting}

To complete the rule, we still need to handle fractions with different denominators.  Our mathematical logic is to convert the original fractions $N_1/D_1$ and $N_2/D_2$ into two fractions each with the common denominator of $D_{sum}=D_1D_2$.  The new numerator of $N_1$ will be $N_1'=N_1D_2$ and that of $N_2$ will be $N_2'=N_2D_1.$ We thus change the problem to that of adding the fractions $N_1D_2/(D_1D_2)$ + $N_2D_1/(D_1D_2)$. Here is this logic implemented in Prolog.

\ps
add(fract(N1,D1),fract(N2,D2),R) :- 
     D1 \= D2, 
     N1p = (N1 * D2),
     N2p = (N2 * D1),
     Dsum = D1 * D2,
     add(fract(N1p,Dsum),fract(N2p,Dsum),R).
\end{lstlisting}

\noindent Again in words, ``add can add two fractions only IF the two denominators are different AND $N_1'$ can be set equal to $N_1D_2$ AND $N_2'$ can be set equal to $N_2D_1$ AND $D_{sum}$ can be set equal to $D_1D_2$ AND the established ``same denominator'' clause can finish the job, placing the result into $R$.'' 

Here we also see more practice on the rules of mathematics: we first form two fractions with the same denominator namely $N_1'/D_{sum}$ and $N_2'/D_{sum}$, then reuse the first clause that knows how to handle like-denominator fraction addition. Both clauses would appear in the Prolog code, and both will be used as needed during the search for solutions.  

Finally, adding two fractions can now be done by specifying this goal using {\tt add}:

\ps
add(fract(3,8),fract(4,7),R).
\end{lstlisting}

\noindent where {\tt R} (for result) is going in uninstantiated (or unassigned), to which Prolog will respond {\tt R = fract(3*7+4*8, 8*7) }. This evaluates to $53/56$ (which is the sum of $3/8+4/7$).

\subsection{Summary}
Using Prolog in the context of mathematics can provide a novel new viewpoint for the student in working with, reviewing, and learning the rules of mathematics and their appropriate use, with the interactivity a computer affords.  We find Prolog's syntactical flexibility makes the mathematical facts+rules implementation an engaging intellectual challenge, which will test one's understanding of the rules themselves.  More examples are presented below.

\section{Non-numerical coding for mathematics}
Looking at math with Prolog is also somewhat contradictory, as Prolog is not known to be a language particularly well suited for the numerical computation. For example, using a for-loop to compute a sum (very natural for in a procedural language), would require a carefully thought out plan involving an accumulator and recursion.~\cite{prolog_count}  Thus, our ideas presented here to study mathematics with Prolog do not focus on having code that generates numerical results (as in the answer to a math problem).   

Prolog is very convenient however, in its ability to manipulate symbols, which may include variables, or objects that might have other meanings such as {\tt fract(1,2)} to represent the fraction $1/2$.  This text and symbolic manipulations is our focus when using Prolog and mathematics.

An an example, in the fraction addition above, the output  was {\tt R = fract(3*7+4*8, 8*7) }.  One may notice that the calculations were not carried out.  This is because the multiplication operator in Prolog or {\tt *} is merely a functor that take two arguments. It is defined in Prolog as {\tt *(A,B)}, similar to our own {\tt fract} used above with its two arguments (a numerator and denominator).  So telling Prolog {\tt 3*7} is merely the infix notation for the functor {\tt *(3,7)}. With this, Prolog does not explicitly see {\tt *} as a directive to multiply, but does has a built in functor {\tt is} that will force a numerical evaluation of some expression (so if one does {\tt P is 3*7}, {\tt P} will indeed hold 21). Thus doing {\tt N1p = N1 * N2} merely instantiates {\tt Np1} to the literal {\tt N1 * N2} (with current values for {\tt N1} and {\tt N2} filled in).

We highlight this point by illustrating two more possibilities of the {\tt add} clause. First, suppose numbers were not even used, as in

\ps
add(fract(red,blue),fract(green,yellow),R).
\end{lstlisting}

This will result in {\tt R = fract(red*yellow+green*blue, blue*yellow).} 

\noindent Second, familiar variables from algebra can also be used as in

\ps
add(fract(x,2*y),fract(2*x,z),R).
\end{lstlisting}

This will result in {\tt R = fract(x*z+2*x* (2*y), 2*y*z)}, which will need to be simplified by the student. 

We find the ``everything is just a symbol'' idea in Prolog to be a desirable property for the study of mathematics, as the meaning of symbols used can be linked to some desired pedagogy.

As these two cases show, Prolog is willing to look at both numbers and symbols as facts, and only treat one or the other differently, as directed by the code.  Because Prolog doesn't immediately differentiate between symbols and possible expressions that could be evaluated, the code unknowingly presents steps for the student to see, and how results came to be.  If Prolog merely output $53/56$ for the fraction addition, the student might still be left confused at how such numbers were found. The sequence of steps toward an answer may be  tracked with an answer like $(3*7+4*8)/(8*7)$, even there is still some work for the student to do (the final simplification). In this context, Prolog may be seen as a ``lazy'' tutor, presenting core logic to the students, but not bothering with the (tedious) details that lead to an actual solution.

There is some contrast here between what Prolog does with these symbols and what would be done by a ``symbolic math'' or computer algebra system (CAS) such as Mathematica~\cite{mma}.  In response to {\tt red/blue + green/yellow}, Mathematica outputs $\frac{red}{blue}+\frac{green}{yellow}$, with no other simplification forthcoming.  In response to {\tt x/(2 y) + 2 x/z}, it outputs $\frac{x}{2y}+\frac{2x}{z}$.  A simplification steps yields $\frac{1}{2}x\left(\frac{1}{y}+\frac{4}{z}\right)$. The {\tt Together} function does handle common denominators as seen in elementary math, but is a ``just the answer'' exercise, without the opportunity to code the needed rules.

\section{Examples}
Here we present three more examples, all of which are available for use online.~\cite{cbm1}. These include solving a right triangle, reasoning about lines and triangles from a collection of points, and converting units.

\subsection{Solving a right triangle}

Consider a Prolog-based example that would help someone solve a right triangle.  The typical exercise would be to present the student with a right triangle, and give them two knowns about the triangle, either two sides, an angle and a side, or two angles.  From this, all angles and sides of the triangle are to be solved for.  

In this example, a model triangle has sides $a$, $b$, and $c$, with $c$ being the hypotenuse, with angles $d$ and $e$, as in Fig.~\ref{rtri}:

\begin{centering}
\begin{figure}[h]
\includegraphics[scale=0.3]{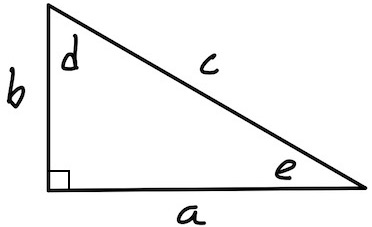}
\caption{A right triangle (see text).}
\label{rtri}
\end{figure}
\end{centering}

The facts of a right triangle are what is known at the onset of the problem, as in ``you are given that $b=5$ and $c=7$.'' As mentioned above, these would be easy to identify, simply by carefully reading the stated problem.  Next, there are eight rules for solving a right triangle: 1) the Pythagorean theorem, 2) the sum of all angles is $180^\circ$, 3-5) the sine, cosine, and tangent of one angle, and 6-8) the sine, cosine and tangent of the other angle. Each of these rules has three inputs, and our logic is that if any two of three are known, the third can be computed using the mathematics of the rule, with the result becoming newly found knowledge, that may trigger the use of another rule. These rules are put into Prolog as shown here:

\ps
rule([c,a,b],'c^2=a^2+b^2').
rule([d,e],'d+e+90=180').
rule([d,a,c],'sin(d)=a/c').
rule([d,b,c],'cos(d)=b/c').
rule([d,a,b],'tan(d)=a/b').
rule([e,b,c],'sin(e)=b/c').
rule([e,a,c],'cos(e)=a/c').
rule([e,b,a],'tan(e)=b/a').
\end{lstlisting}

Each rule contains a list of required variables, followed by a (minimal) description of a right triangle rules that relates the variables.  The goal, not shown here (but presented online~\cite{cbm1}) is called {\tt advise}, and is passed in a list of triangle quantities that are known (the facts), to start Prolog's search.  Running for example {\tt advise([b,c]).} would result in the following output:

\ps
Solve for a using 
c^2=a^2+b^2
You now know 
[a,b,c]

Solve for d using 
sin(d)=a/c
You now know 
[d,a,b,c]

Solve for e using 
d+e+90=180
You now know 
[e,d,a,b,c]
\end{lstlisting} 

Once again, the emphasis is not to generate numerical results. The ``solution'' is logical method of solving the triangle that will lead to the needed results, while leaving any numerical computation up to the student.

\subsection{Reasoning about a collection of points}

In this example, the facts are a collection of points placed on a standard $xy$-coordinate grid, as shown here:

\ps
point(-4,4).
point(-3,2).
point(-1,1).
point(0,0).
point(-3,0).
point(1,3).
point(3,3).
point(-4,-4).
point(3,-3).
\end{lstlisting}

\noindent With these points, we implement some Prolog code to ``reason'' about them, in particular, find lines made by pairs of points, then lines that are horizontal and vertical.

As we know, a line is made of any two points, so Prolog code that will search for such lines amongst the facts (of points) would look like this:

\ps
line([(X1,Y1),(X2,Y2)]) :- 	
     point(X1,Y1), 
     point(X2,Y2), 
     point(X1,Y1) \= point(X2,Y2).
\end{lstlisting}

Here we see a clause called {\tt line} that is associated with a list of two $xy$-points. (A list in Prolog is a comma separated sequence of objects enclosed in {\tt [} and {\tt ]}.) The clause would read like this: ``A line between points {\tt (X1,Y1)} and {\tt (X2,Y2)} exists IF a point at {\tt (X1,Y1)} exists AND a point at {\tt (X2,Y2)} exists, AND be sure that the two points are different (to ensure we don't have line whose endpoints are both the same point). This goal run against the point facts given will yield:

\ps
L = [(-4,4),(-3,2)] ;
L = [(-4,4),(-1,1)] ;
L = [(-4,4),(0,0)] ;
L = [(-4,4),(-3,0)] ;
L = [(-4,4),(1,3)] ;
L = [(-4,4),(3,3)] ;
L = [(-4,4),(3 , -3)] ;
...
...
L = [(3 , -3),(1,3)] ;
L = [(3 , -3),(3,3)] ;
L = [(3 , -3),(-4 , -4)] ;
\end{lstlisting}

With lines now established, we can also check if a line is horizontal with this rule:

\ps
horiz([(_,Y),(_,Y)]).
\end{lstlisting}

This short rule says a line is horizontal if both $y$-coordinates are the same.  Since the $x$-coordinate is immaterial, we put in Prolog's ``anonymous'' or ``don't care'' variable in for the $x$-coodinates, which is an underbar ({\tt \_}). This clause would also work: {\tt horiz([(\_,Y1),(\_,Y2)]) :- Y1 = Y2.}  If variables were put in for $x_1$ and $x_2$ (in either clause), as in {\tt horiz([(X1,Y1),(X2,Y2)])}, Prolog would throw a warning since they are not used in the definition of a horizontal line. Testing for vertical lines is similar, as in {\tt vert([(X,\_),(X,\_)]).}.

Now we may look for horizontal lines amongst the points.  The goal for this would be {\tt line(L), horiz(L).}  Here we first find a line $L$ amongst the points, then test if $L$ is horizontal. For the point data shown, Prolog yields:

\ps
L = [(0,0),(-3,0)] ;
L = [(-3,0),(0,0)] ;
L = [(1,3),(3,3)] ;
L = [(3,3),(1,3)] ;
\end{lstlisting}

The student may verify (by hand) that these are indeed horizontal lines.  Vertical lines can be found using the goal {\tt line(L), vert(L).} and will output:

\ps
L = [(-4,4),(-4 , -4)] ;
L = [(-3,2),(-3,0)] ;
L = [(-3,0),(-3,2)] ;
L = [(3,3),(3 , -3)] ;
L = [(-4 , -4),(-4,4)] ;
L = [(3 , -3),(3,3)] ;
\end{lstlisting}

From horizontal and vertical lines, we can also find perpendicular lines, which have two considerations.  The first is having two lines $L_1$ and $L_2$, with one is vertical and the other is horizontal.  This requires two clauses as per

\ps
perp(L1,L2) :- horiz(L1), vert(L2).
perp(L1,L2) :- vert(L1), horiz(L2).
\end{lstlisting}

This read ``Lines $L_1$ and $L_2$ are perpendicular if one is horizontal and one is vertical (or vice-versa).'' (We note here the lines are thought of as containing the $(x,y)$ points in the data, but not necessarily as their end-points.) Presenting Prolog with the goal of {\tt line(A), line(B), perp(A,B)} will generate this (partial) output:

\ps
A = [(0,0),(-3,0)], B = [(-4,4),(-4 , -4)] ;
A = [(0,0),(-3,0)], B = [(-3,2),(-3,0)] ;
A = [(0,0),(-3,0)], B = [(-3,0),(-3,2)] ;
A = [(0,0),(-3,0)], B = [(3,3),(3 , -3)] ;
A = [(0,0),(-3,0)], B = [(-4 , -4),(-4,4)] ;
A = [(0,0),(-3,0)], B = [(3 , -3),(3,3)] ;
A = [(-3,0),(0,0)], B = [(-4,4),(-4 , -4)] ;
...
...
\end{lstlisting}  
The second consideration shown online~\cite{cbm1} considers two lines, neither of which is vertical, whose slopes have a product of $-1$.

Online, the lessons are extended to include looking for right triangles within the points via

\ps
line([P1,P2]), 
line([P2,P3]), 
line([P3,P1]), 
perp([P1,P2],[P2,P3]).
\end{lstlisting}

This clause supposes that a triangle exists amongst the points $P_1$, $P_2$ and $P_3$, and starts by looking for three lines containing these points. Note the usage of these points in the search for the three lines that will make a (closed) triangle, in particular that $P_1$ connects to $P_2$, $P_2$ to $P_3$ and $P_3$ back to $P_1$.  Once suitable lines are found, at least two of them are checked for perpendicularity.  Rectangles can be found similarly, by looking for 4 lines and 3 pairs that are perpendicular. Circles could be found using a rule that looks for points that are all equidistant from a center point.

\subsection{Converting Units}

Converting units includes problems like ``convert 10 inches in to kilometers.'' Initial lessons usually require a strict accounting of the units themselves, to be sure they all cancel properly.  An example is

\begin{equation*}
\frac{10 \cancel{\textrm{ inches}}}{1}\times\frac{2.54 \cancel{\textrm{ cm}}}{1 \cancel{\textrm{ inch}}}\times\frac{1 \cancel{\textrm{ m}}}{100 \cancel{\textrm{ cm}}}\times\frac{1 \textrm{ km}}{1000 \cancel{\textrm{ m}}},
\end{equation*}

which requires the student to find a step-by-path from the original unit to the desired (or terminating) unit.  The facts for such a problem would be easily found, many of which are commonly known ways of directly converting one unit into another, such as this collection:

\ps
fact(1,ft,12,inch).
fact(1,yard,3,ft).
fact(1,mile,5280,ft).
fact(220,yard,1,furlong).
fact(2.54,cm,1,inch).
fact(10,mm,1,cm).
fact(100,cm,1,meter).
fact(1000,meter,1,km).
\end{lstlisting}

The issue at hand is in searching the facts that will find a path from the original unit to the desired final unit. For this discussion, we'll focus on converting inches into kilometers (km).  Using only the facts shown, a quick scan would see the steps 1) inches to cm 2) cm to meter, them 3) meter to km.  Algorithmically, this is not an easy task, either in a declarative or procedural language.  In Prolog, the facts are straightforward handle, as shown above.  Our solution proceeds as follows.

To begin, we expect the facts to work in both directions. That is, if a fact states how to convert from unit $A$ to $B$, this same fact can be used to convert from $B$ to $A$.  This is not automatically handled by Prolog, so we include a rule called {\tt direct} (as in ``direct conversion'') as follows:~\cite{wontwork}

\ps
direct(X, Y) :- fact(_,X,_,Y).
direct(X, Y) :- fact(_,Y,_,X).
\end{lstlisting}

Note the reversal of $X$ and $Y$ in the 2nd clause.  When looking for conversion facts, the code will always used {\tt direct} and not {\tt fact}.

The top-level conversion rule is {\tt convert(From,To,How)}, which will convert from a unit {\tt From} to a unit {\tt To}, with directions for how to do so (given the facts) returned in {\tt How}.  

This rule calls another rule called {\tt path}, which treats the facts given as a tree to traverse, looking for a path from unit {\tt From} to unit {\tt To}.  By ``tree,'' we mean each unit in a fact is a node that has a variable number of branches from it, to units it can be directly converted into. This is shown in Fig.~\ref{units_tree}.

\begin{centering}
\begin{figure}[h]
\includegraphics[scale=0.2]{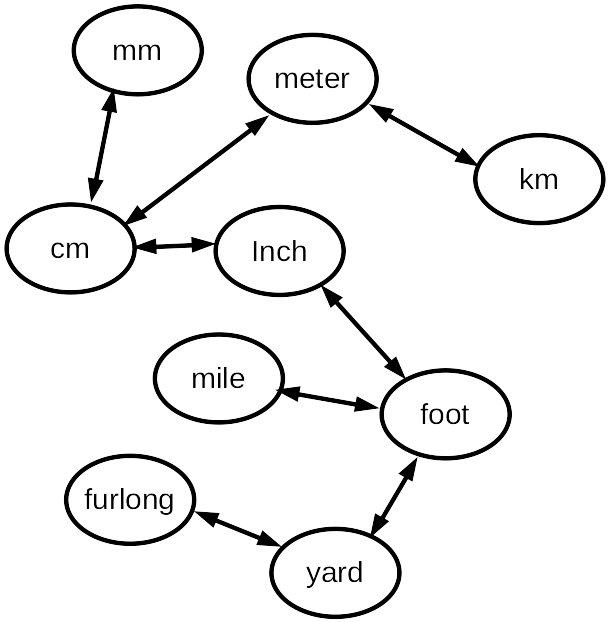}
\caption{A tree representation of the unit conversion facts presented.}
\label{units_tree}
\end{figure}
\end{centering}

The algorithm's job is to find a path through the tree from the original unit to the terminating unit.  The rule {\tt path} is complicated, as was adapted from a treatment by Clocksin~\cite{clocksin} on searching graphs. We explain it here.

The core logic here is that if unit $X$  cannot be directly converted into $Y$ from the stated facts, we'll have Prolog search for what $X$ {\sl can} be converted into, in this case some intermediate unit $Z$.  Then we see if $Z$ can be converted into $Y$.  Thus the logic is to find a {\tt direct} from $X$ to $Z$ as in {\tt direct(X,Z).} We must keep track of what we have converted to in a variable {\tt Trail} as to avoid endless and repeated conversions into the same units.  (Like the need for the {\tt direct} clause, this is also something Prolog does not handle automatically.)  Lastly, once we find a unit $Z$ that $X$ can be converted into, we re-enter the same search in a recursive manner, now looking to convert $Z$ to $Y$. (This may lead to finding an intermediate unit that $Z$ can be converted to, and so on; this is the power of recursion.)  This is all handled in the {\sl recursive} clause of {\tt path}:

\ps
path(X,Y,Trail,How0) :- 
     direct(X,Z), 
     \+ member(Z,Trail),  
     path(Z,Y,[X|Trail],How0).
\end{lstlisting}

The variables are {\tt X} and {\tt Y}, the starting and terminal units, {\tt Z} is some intermediate unit, {\tt Trail} keeps track of where the search has gone to avoid repeats, and {\tt How0} is the directions that will be given as output.  With this, the rule reads ``it is possible to find a path from {\tt X} to {\tt Y} keeping track in {\tt Trail} and directions in {\tt How0} IF the facts allow for direct conversion of {\tt X} into some unit {\tt Z} AND {\tt Z} is not ({\tt $\backslash$+}) already in the {\tt Trail} AND a path from {\tt Z} to {\tt Y} can be found, while prepending {\tt X} onto {\tt Trail} and keeping {\tt How0} available for final directions.''

Since this is a recursive call, we must have a terminal condition. This is another rule, which is a {\sl non-recursive} clause for {\tt path} which reads:

\ps
path(X,X,Trail,[X|Trail]).
\end{lstlisting}

This terminates the recursion if we try to find a path from unit {\tt X} to the same unit {\tt X}.  This is an odd expectation, going from {\tt X} to {\tt X}, as why would Prolog try to do this?  Also, we'd never expect a fact to be given that shows us how to go from unit to the same unit.  We use this oddity as an opportunity to illustrate the execution order of Prolog.  

Those familiar with recursion in procedural languages might think the {\tt path} call in the recursive clause will again reenter the recursive clause. Prolog will not do this, as it always calls rules in the order in which they appear in the code. Here the non-recursive terminal clause appears before the recursive one, so the terminal rule will always be the first called when {\tt path(...)} is ever queried.

It is still a curious point on when and why the unit path search would try to convert from  one unit to the same unit, terminating the recursion. This happens when the intermediate unit {\tt Z} reaches the desired terminal unit.  One notes in the recursive rule

\ps
path(X,Y,Trail,How0) :- 
     .
     .  
     path(Z,Y,[X|Trail],How0).
\end{lstlisting}

\noindent that the parameter {\tt Y} is always held as the second argument to {\tt path}.  Thus in all of the recursive calls, the desired terminal unit is always preserved in variable {\tt Y}, in both the non-recursive (terminal) and recursive clauses. This is how the terminal clause is triggered: when the intermediate unit {\tt Z} finally matches the desired terminal unit, which is always held in this second argument position.~\cite{hard}

If we call on convert then, with something like {\tt convert(inch,km,Plan).}, Prolog will respond with {\tt Plan = [inch,cm,meter,km]}, which indeed it a step by step conversion from inches to km, using the  direct conversions possible, as supplied by the facts. The entire code is available for running live online.~\cite{cbm1}

\subsection{Prolog in use}

The advantage to using Prolog versus a procedural language here, as it ties to a quote in Ref.~\citenum{van_hetenryck}, is somewhat clear. In the examples presented,  reflect how continually the meaning of the needed rules are in plain view, and how straightforward it is to change or extend them.  In the example, the logic of perpendicular lines is always there, with the actual code more or less saying ``perpendicular lines A and B exist if A is vertical and B is horizontal or vice-versa.''  Facts are in a similar situation: adding or editing known units conversion facts are a few keystrokes away, as are adding or changing $(x,y)$ points.

\section{Suggestions on using Prolog}

We close this paper by sharing a few of our own observations and experiences using Prolog.  We have been using Prolog on and off since the late 1980s, and have managed to produce some published work using it.~\cite{bensky} We always think of Prolog as a fun, powerful and intriguing tool for solving problems with code, but also find it perennially difficult use.~\cite{difficult_why}  The mathematical context discussed in this paper is an actual a learning motivator for us.

As for learning, many books on Prolog were written in the 1980s, and are now out of print. It is rare to find one on the shelf of a bookstore (even used). Some are now free and in the public domain.~\cite{covington_collection}. Few (if any) new books have appeared since, although two of the best books on the subject (that we hold in high regard), have been revised somewhat recently.~\cite{clocksin,bratko}. Online resources like ``Learn Prolog now''~\cite{lpn}, the ``Power of Prolog,''~\cite{pop} and other collections of lessons~\cite{fisher} are very instructive.  

Adding to Prolog's difficulty is a lack a modern, dedicated, and true IDE, such as what the Spyder IDE does for Python.~\cite{spyder}  Most work we do with Prolog is (painfully) at the Unix prompt. If one wishes to lightly experiment with Prolog, we'd recommend finding an online IDE~\cite{swish,cbm_sandbox} to get started.

Often Prolog code found in books and online lessons comes in the form of challenging puzzles such as the N-queens problem, traveling salesperson, or solving a Sudoku. While these leave one with a ``gee-whiz'' feeling of what Prolog can  do, they give very little inspiration for one's everyday coding needs.  Practical uses for Prolog seem hard to come by, but we again refer to Ref.~\citenum{difficult_why}. Common introductory lessons on Prolog are important, yet uninspiring, as in ``write a clause to see if item A is a member of list L,'' which is a very common example. This can be contrasted with tutorials on procedural languages, that may quickly get into graphics, web-apps, and the like.

State-of-the-art Prolog compilers and interpreters do continue to see regular development~\cite{swi,gprolog,tau}, and constraint logic programming (CLP) seems to be reinvigorating the language.~\cite{trista}

CLP has the following power in the declarative nature of Prolog:  Suppose a clause you are working on says {\tt X = Y +1}.  If {\tt Y} is not known, the clause will fail, since {\tt X} cannot be instantiated. With CLP, this would be represented with {\tt X \#= Y +1}, where {\tt \#=} is the CLP version of {\tt =}. In this case, evaluation of the clause will continue, where Prolog will consider satisfying {\tt X \#= Y+1} either by finding {\tt Y}, in which case {\tt X} can be found,  or {\tt X}, in which case {\tt Y} can be found.  A first-order understanding of this tells us that CLP allows for elementary algebra to work within a clause, which indeed is a powerful mechanism to add to Prolog's search mechanisms. A higher-order thought is that CLP kind of allows Prolog to run code in reverse, as in telling you what inputs will produce some output you may desire.

We see applications of CLP in studying math with Prolog. Returning to the fraction addition, {\tt add(fract(N1,D1),fract(N2,D2),R)}, CLP would allow us to pose a goal of {\tt add(fract(2,D1),fract(3,7),frac(1,2))} for example.   This would have Prolog search for the denominator {\tt D1} of $2/D_1$ such that $2/D_1+3/7=1/2$.  CLP will also work with more than one unknown variable. Expected ranges of solutions can be specified to narrow the search, but even without such, CLP over the integer domain comprises a finite search field.  Another interesting example of using CLP in mathematics is in dealing with Fibonacci numbers in reverse.~\cite{fib}

Lastly, in a production environment, we don't see Prolog as well suited to be the primary code base. It has always had weak input and output capabilities, and sanitizing input data from a user in Prolog would simply not be worth the effort.  Instead, we see a front-end procedural language used to generate Prolog code based on user input, perhaps as assembling a bunch of facts to be paired with some pre-written rules.  The procedural language would then run a Prolog environment, piping in the generated code and capturing the results coming out, which can then be formatted and presented to the user.

\section{Conclusions}

We have shown how Prolog can be used to examine topics in elementary mathematics. We find the declarative nature of the language well suited for this task, even more so with CLP. Prolog may provide a potential platform for creating tutorials, lessons, and student projects involving mathematics.  Examples presented from elementary math, including addition of fractions, reasoning about $(x,y)$ points, and conversion of units have been presented, with more found online.~\cite{cbm}


\begin{thebibliography}{99}
\bibitem{engel} A. Engel, ``Exploring Mathematics with your Computer (Anneli Lax New Mathematical Library, Series Number 35),'' The Mathematical Association of America; Pap/Dskt edition (August 14, 1997).
\bibitem{cheney} E.W. Cheney and D.R. Kincaid, ``Numerical Mathematics and Computing,'' (7th ed.), Cengage Learning, (2012).
\bibitem{becker} K. Becker and Michael Dorfler, {\sl Dynamical Systems and Fractals, Computer Graphics Experiments with Pascal}, Cambridge, 1989.
\bibitem{devaney} R.L. Devaney, {\sl Chaos, Fractals, and Dynamics}, Addison-Wesley, 1990.
\bibitem{guggenheimer} H. Guggenheimer, {\sl BASIC Mathematical Programs for Scientists and Engineers}, Petrocelli, 1987.
\bibitem{wolframbook} S. Wolfram, ``An Elementary Introduction to the Wolfram Language - Second Edition
An Elementary Introduction to the Wolfram Language - Second Edition,'' Wolfram Media, 2017.
\bibitem{conrad} C. Wolfram, ``The Math(s) Fix: An Education Blueprint for the AI Age,'' Wolfram Media, 2020.
\bibitem{websites} See https://projecteuler.net and https://codebymath.com.
\bibitem{clocksin} W. Clocksin and C. Mellish, ``Programming in Prolog,'' (5th ed.), Springer, 2003.
\bibitem{bratko} I. Bratko, ``Prolog Programming for Artificial Intelligence,'' (3rd ed.), Addison-Wesley, 2001.
\bibitem{covington} M.A. Covington, D. Nute, and Andre Vellino, ``Prolog Programming in Depth,'' (2nd ed), Prentice-Hall, 1997.
\bibitem{bos} J. Bos, ``Drawing Prolog Search Trees: A Manual for Teachers and Students of Logic Programming,'' {\tt arXiv:2001.08133 [cs.PL]}.
\bibitem{van_hetenryck} P.V. Hentenryck, ``Constraint Satisfaction in Logic Programming,'' MIT Press, 1989. This author has some compelling comments on translating seach-type algorithms to procedural languages: ``They achieve a reasonable efficiency but are tedious to develop, extend and maintain.  The amount of programming is generally very significant and simple conceptual changes in the problem specification or in the resolution strategy cannot be accommodated easily.''
\bibitem{cbm} This example can be run and used here \url{https://www.codebymath.com/index.php/welcome/lesson/prolog-add-fractions}.
\bibitem{pop_reading} See ``Reading Prolog Programs'' at \url{https://www.metalevel.at/prolog/reading}.
\bibitem{prolog_count} See \url{https://www.codebymath.com/index.php/welcome/lesson/prolog-count}.
\bibitem{mma} See Mathematica at \url{https://wolfram.com}.
\bibitem{cbm1} See \url{https://www.codebymath.com}.
\bibitem{wontwork} We note that {\tt fact(X,Y) :- fact(Y,X).} will not work. This instead leads to a recursion overflow, since {\tt fact} continually calls itself, with no opportunity to exit.
\bibitem{hard} We find this search algorithm clever and instructive but was not very intuitive to us.  It illustrates the difficulty one must consider when handling looping and accumulators in Prolog, but also note how short (and still declarative) the code is, as compared to an equivalent implementation in a procedural language. Such complications are likely why Prolog has never seen widespread adoptions: looping is always based on recursion, which is an advanced and relatively difficult programming topic versus for example, a basic for-loop found in procedural languages.
\bibitem{bensky} T.J. Bensky and C.A. Taff, ``Computer-Guided Solutions to Physics Problems Using Prolog,'' {\sl Comp. Sci. Eng.}, Vol. 12, Iss. 1, Jan. 1, 2010, pages 88-95. The paper can be found here: \url{https://github.com/tbensky/physics\_prolog/blob/master/paper.pdf}.

\bibitem{difficult_why} There are reasons for this.  Prolog is a powerful language. Its free and expressive format opens up one's thinking on some of the strangest coding problems we might consider.  Scheduling? Understanding natural language? Searching a graph? Advising a student on how to learn math? In this work alone, we presented short, digestible code that gives us step-by-step directions on how solve a right triangle!  In any procedural language, this would be loaded with arrays, loops, and calls to string processing libraries.  The result would be a boring ``so what'' kind of outcome not really worth presenting.  In Prolog we have a list of variables directly paired with mathematical expressions at the onset. This is about as low of a thought-to-code barrier as one could possibly expect.  Although not always easy either, we simply don't use Prolog for some task like opening a file, reading data from columns \#2 and \#7 to making an XY plot. We'd use Python for this. So we think the difficulty isn't really with Prolog per se, {\sl but with the problems Prolog allows us to think about solving}. They are difficult problems anyway, and would also be made even more so if solved in a procedural language.  

\bibitem{covington_collection} See \url{http://www.covingtoninnovations.com/books.html}.
\bibitem{lpn} See \url{http://www.let.rug.nl/bos/lpn//index.php}.
\bibitem{pop} See ``Power of Prolog'' at \url{https://www.metalevel.at/prolog/}.
\bibitem{fisher} See \url{https://skolemmachines.org/ThePrologTutorial/www/prolog_tutorial/contents.html}.
\bibitem{spyder} Imagine something like the Spyder IDE at \url{https://www.spyder-ide.org}, which is (for Python), but for Prolog.
\bibitem{swish} See \url{https://swish.swi-prolog.org.}

\bibitem{cbm_sandbox} See \url{https://www.codebymath.com/index.php/welcome/sandbox}. Place the line {\tt \% prolog} as the first line of code to trigger the integrated Prolog interpreter.~\cite{tau}
\bibitem{swi} See \url{https://www.swi-prolog.org}.
\bibitem{gprolog} See \url{http://www.gprolog.org}.
\bibitem{tau} See \url{http://tau-prolog.org}.
\bibitem{trista} See ``The Power of Prolog'' at  \url{https://www.metalevel.at/prolog}, and in particular \url{https://www.metalevel.at/prolog/optimization}. We note that both SWI and Gnu-Prolog support constraint logic programming.
\bibitem{fib} See \url{https://m00nlight.github.io/constraint%20logic%20programming/2016/01/01/using-clpfd-to-solve-factorial-and-fibonacci-problems-reversely}.


\end{thebibliography}
\end{document}